\documentclass{osa-article}

\journal{oe}

\articletype{Research Article}

\begin{document}

\title{Untrained networks for compressive \\ lensless photography}

\author{Kristina Monakhova,\authormark{1, $\dagger$, *} Vi Tran,\authormark{2, $\dagger$} Grace Kuo,\authormark{1} and Laura Waller\authormark{1}}

\address{\authormark{1}Department of Electrical Engineering \& Computer Sciences, University of California, Berkeley, CA, 94720, USA\\
\authormark{2}Orange Coast College, Costa Mesa, CA,92626, USA\\
\authormark{$\dagger$ These authors contributed equally to this work}}

\email{\authormark{*}monakhova@berkeley.edu} 



\begin{abstract}
Compressive lensless imagers enable novel applications in an extremely compact device, requiring only a phase or amplitude mask placed close to the sensor. They have been demonstrated for 2D and 3D microscopy, single-shot video, and single-shot hyperspectral imaging; in each case, a compressive-sensing-based inverse problem is solved in order to recover a 3D data-cube from a 2D measurement. Typically, this is accomplished using convex optimization and hand-picked priors. Alternatively, deep learning-based reconstruction methods offer the promise of better priors, but require many thousands of ground truth training pairs, which can be difficult or impossible to acquire. In this work, we propose an unsupervised approach based on untrained networks for compressive image recovery.  Our approach does not require any labeled training data, but instead uses the measurement itself to update the network weights. We demonstrate our untrained approach on lensless compressive 2D imaging, single-shot high-speed video recovery using the camera's rolling shutter, and single-shot hyperspectral imaging. We provide simulation and experimental verification, showing that our method results in improved image quality over existing methods.
\end{abstract}

\section{Introduction}
Compressive imagers aim to recover more samples than they measure by leveraging compressive sensing, which guarantees signal recovery for underdetermined systems given certain assumptions about signal sparsity as well as incoherence in the measurement domain~\cite{candes2006compressive, candes2008introduction}. For optical imaging, compressive imagers have been demonstrated for many applications, including single-pixel and coded-aperture cameras~\cite{duarte2008single, wakin2006architecture, huang2013lensless}. These imagers have been shown to be effective for a number of compressive sensing tasks, such as single-shot lightfield imaging~\cite{marwah2013compressive}, 3D imaging~\cite{levin2007image}, hyperspectral~\cite{wagadarikar2008single, gehm2007single}, and high-speed video imaging~\cite{hitomi2011video, reddy2011p2c2, llull2013coded, gao2014single}. Recently, there has been a push to make compressive imagers more accessible by using inexpensive and compact hardware~\cite{jeon2019compact, liutkus2014imaging}. One such promising method includes lensless, mask-based imagers, which remove the lens of a traditional camera and instead place a phase or amplitude mask near the sensor to randomize the measurement, approximately achieving the measurement domain incoherence required for compressive sensing. These lensless mask-based cameras can be incredibly compact, with the mask only millimeters from the sensor, and assembly does not require precise alignment~\cite{asif2016flatcam, antipa2018diffusercam, shimano2018lensless}. Mask-based lensless imagers have been demonstrated for compact single-shot 3D fluorescence microscopy~\cite{adams2017single, kuo2020chip, yanny2020miniscope3d, liu2020fourier}, hyperspectral imaging~\cite{monakhova2020spectral}, and high speed single-shot video~\cite{antipa2019video}. These sorts of cameras are very promising for a number of imaging modalities; however, their performance depends on the fidelity of the reconstruction algorithm. Since the algorithm must solve an underdetermined inverse problem with assumptions on signal sparsity, the choice of algorithm and imaging priors used can significantly impact results. 

\begin{figure}[h]
\includegraphics[width=\linewidth]{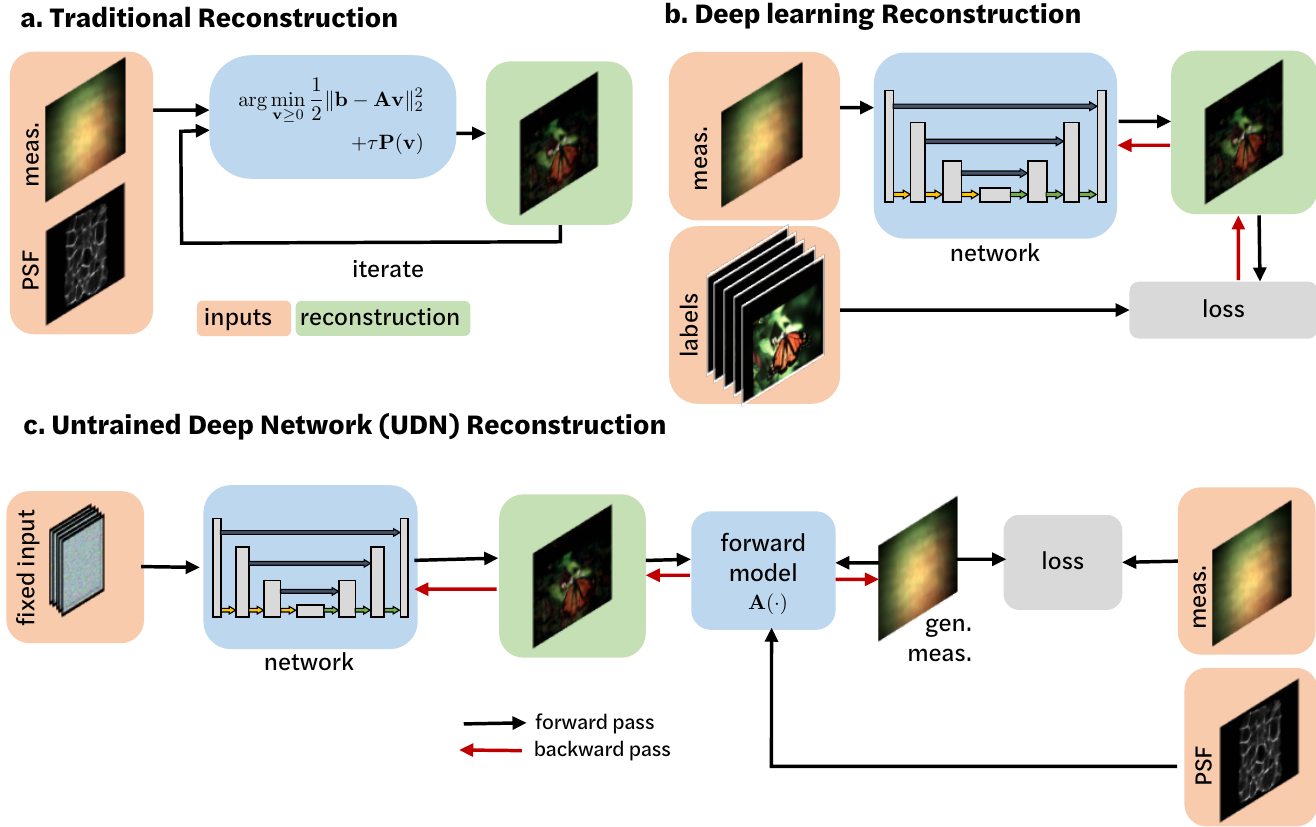}
 \caption{\textbf{Comparison of different reconstruction approaches}. (a)~Traditional reconstructions solve a convex optimization problem with a data-fidelity-term based on the known forward model and a regularization term which acts as the image prior.  (b)~Deep learning methods require thousands or more labeled image pairs in order to train a neural network to approximate the reconstruction.  (c)~Our untrained deep network (UDN) uses a neural network as the image prior, but does not require any training data.  The system forward model is used to calculate the loss between the estimated image and the measurement, which is then used to update the network weights.}
\label{fig:comparison}
\end{figure}

Traditionally, images from compressive cameras are recovered by solving a convex optimization problem, minimizing both a least-squares loss based on the physics of the imaging system and a hand-chosen prior term, which enforces sparsity in some domain, Fig.~\ref{fig:comparison}(a). For successful image recovery by compressive sensing, there must be both incoherence in the sensing basis and sufficient sparsity in the sample domain. Lensless mask-based cameras utilize multiplexing optics to fulfill the incoherent sampling requirement, mapping each pixel in the scene to many sensor pixels in a pseudo-random way.  The prior term enforces sparsity and ensures successful signal recovery. Over the years, a number of different hand-picked priors have been used for compressive imaging, such as sparsity in wavelets, total-variation (TV), and learned dictionaries. TV, which is based on gradient sparsity, has been particularly popular~\cite{li2010efficient}. These methods are effective at solving imaging inverse problems, however they have recently been outperformed by deep learning-based methods, which incorporate non-linearity and network structures that may be better suited to image representation~\cite{jin2017deep}.  Recent work on plug and play (PnP) priors~\cite{ryu2019plug, venkatakrishnan2013plug, zhang2020plug, romano2017little} offers some hope of combining inverse problems with state of the art denoisers (both deep and not, e.g. BM3D~\cite{dabov2006image}); however, PnP still relies on using hand-picked denoisers or pre-trained networks that may not be well-suited for a given application.

Recently, several deep-learning based approaches have shown improved reconstruction quality for imaging inverse problems~\cite{jin2017deep, sinha2017lensless, mccann2017convolutional}; however, they rely on having a large dataset of experimental labeled ground truth and measurement pairs.  In these methods, a deep neural network is used to approximate the imaging inverse problem, Fig.~\ref{fig:comparison}(b). The network is typically trained end-to-end and requires large datasets of labeled image and ground truth pairs. The network can be agnostic to the imaging system physics, or the network can incorporate knowledge of the physics for faster convergence and reduced training data requirements~\cite{monakhova2019learned, khan2020flatnet}. For high-dimensional imaging applications, such as 3D and hyperspectral lensless imaging, obtaining ground truth datasets can be impractical, costly, or impossible.  While there has been some work in training networks from synthetic data, these methods can suffer from model-mismatch if the synthetic data does not match the experimental system~\cite{zhang2019zoom}.  

Recent work using unsupervised learning with untrained networks is especially promising for a number of imaging applications - leveraging the structure of neural networks without needing any training data.  Untrained networks, such as deep image prior~\cite{ulyanov2018deep} and deep decoder~\cite{heckel2018deep}, have shown that the structure of neural networks can be effective at serving as a prior on image statistics without any training. An untrained deep network with randomly initialized weights is used as an image generator, outputting the recovered image.  The network weights are updated through a loss function comparing the generated image with the input data, for example, a noisy image. This method and several related papers have been shown to be particularly effective for simulated image denoising, deblurring, and super-resolution~\cite{mataev2019deepred, liu2019image}. For many computational imaging problems, the measurements do not resemble the reconstructed image; instead, the scene is related to the measurement through a forward model that describes the physics of the image formation problem. For this class of problems, untrained networks can be paired with differentiable imaging forward models in which the network weights are updated based on a loss between the measurement and the generated measurement from the network output passed through the imaging forward model, Fig.~\ref{fig:comparison}(c). This has been shown to be effective on several imaging modalities, such as phase imaging~\cite{bostan2020deep}, MRI~\cite{gong2018pet,cui2019pet, jin2019time}, diffraction tomography~\cite{zhou2020diffraction}, and several small-scale compressive sensing problems~\cite{van2018compressed}. Here, we extend this framework to compressive lensless photography.

We propose a reconstruction method for compressive lensless imaging based on untrained networks, which we call untrained deep network (UDN) reconstructions. Our approach uses a deep network for the image prior, but requires no training data. We present a general differentiable imaging model that could be used for multiple types of compressive lensless imaging, and test it on three different cases: 2D lensless imaging with erasures, single-shot video, and single-shot hyperspectral imaging. 

To the best of our knowledge, untrained networks have not been used for compressive optical imaging before, thus providing both a stress test of untrained deep networks on several challenging underdetermined experimental systems, as well as bringing improved image quality to compressive lensless imaging. We provide simulation and experimental results, showing improved performance over existing methods that do not utilize training data. Our results indicate that untrained networks can serve as effective image priors for these systems, providing better reconstructions in cases where it is not possible to obtain labeled ground truth data.

\section{Imaging models}
\begin{figure}[tp]
\includegraphics[width=\linewidth]{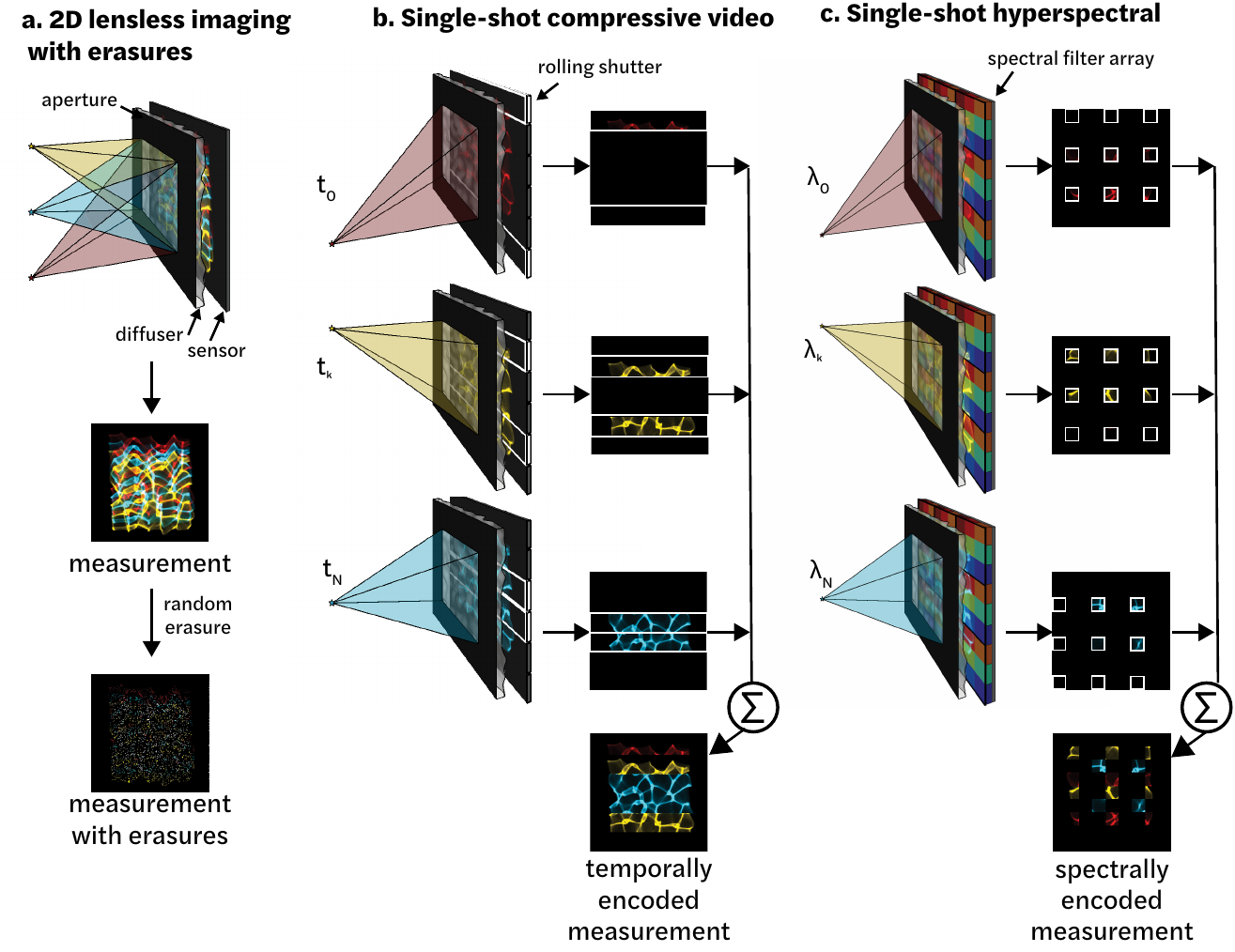}
   \caption{\textbf{Imaging systems:} Our lensless 2D camera consists of a diffuser placed a small distance away from the sensor. Each point in the world maps to a high-contrast caustic pattern (PSF). (a) Since the PSF covers a large area of the sensor, we can erase a random subset of the pixels and still recover the full image. (b) The camera's rolling shutter can be used to encode temporal information into the measurement. As points flick on/off, the PSFs are filtered by the sensor's rolling shutter function, which reads one row at a time (or two in the case of dual shutter, as shown here). The measurement is a sum of the rows captured at different time points, each of which contains information from the entire 2D scene. Hence, temporal information is encoded vertically across the sensor, with earlier time points at the top and bottom of the image and later time points towards the center. (c)  Single-shot hyperspectral imaging is achieved with a spectral filter array in front of the sensor, which maps each band of wavelengths to a subset of the sensor pixels. }
\label{fig:forward}
\end{figure}

Our system architecture is based on the lensless mask-based camera, DiffuserCam~\cite{kuo2017diffusercam, antipa2018diffusercam}, which consists of a thin diffuser (a smooth pseudorandom phase optic) which is placed a few millimeters in front of the sensor, Fig.~\ref{fig:forward}(a).  Light from each point in the scene refracts through the diffuser to create a high-contrast pseudorandom caustic pattern on the sensor plane.
Since each point in the world maps to many pixels on the sensor, we need only a subset of the sensor pixels to reconstruct the full 2D image, given sufficient sparsity. This can be useful, for example, to compensate for dead sensor pixels, which act like an erasure pattern; or, we can capture the full 2D sensor image and use compressed sensing to recover higher-dimensional information. The higher-dimensional data (e.g. time, wavelength) must be physically encoded into different subsets of the sensor pixels; then compressed sensing is applied to recover a 3D datacube.

We demonstrate our UDN on two examples of three-dimensional data recovery from a single-shot 2D measurement: (1) high-speed video~\cite{antipa2019video} and (2) hyperspectral imaging~\cite{monakhova2020spectral}. The high-speed video example takes advantage of the sensor's built-in rolling shutter, which exposes different rows of pixels at each time point, Fig.~\ref{fig:forward}(b), effectively acting like an erasure pattern at each time point. From this data, one can use compressed sensing to recover a separate 2D image for each time point; the result is a full-resolution video at the framerate of the rolling shutter (the line scan rate). For the hyperspectral imaging example, the DiffuserCam device has an added spectral filter array in front of the sensor, with 64 different color channels, Fig.~\ref{fig:forward}(b). Each of the 64 wavelengths thus map to a different subset of the sensor pixels, similarly acting like an erasure pattern for each wavelength. From this data, one can use compressed sensing to recover a hyperspectral datacube (2D spatial + 1D spectral) from a single 2D measurement. In both examples, we choose to reconstruct the full datacube simultaneously, rather than a sequential set of 2D reconstructions, since this allows us to add priors along the time/wavelength dimension.

We now go into detail on the modeling, starting with the 2D image formation model of DiffuserCam, and building to our three example scenarios: 2D imaging with erasures,  high-speed video, and hyperspectral imaging. These formulations will be used to create the differential forward model within our network architecture.

\subsection{2D lensless imaging}

As described in~\cite{kuo2017diffusercam, antipa2018diffusercam}, we model our lensless camera as shift-invariant: we assume the pattern on the sensor from each point in the world is a translated version of the on-axis pattern, called the point spread function (PSF). We can  therefore model the sensor response $\mathbf{b}[x,y]$ as a convolution of the scene $\mathbf v[x,y]$ with an on-axis PSF $\mathbf h[x,y]$:

\begin{align}
\label{eq:forward_2D}
 \mathbf{b}[x,y] &= \text{crop} \Big( \mathbf v[x,y] * \mathbf h[x,y]) \Big)\\
 &= \mathbf A_{\text{2D}} \mathbf v.
\end{align}

\noindent where $[x,y]$ represents the discrete image coordinates, $*$ represents a discrete 2D linear convolution and the crop function accounts for the finite sensor size. For 2D imaging, we assume that objects are placed beyond the hyperfocal distance of the imager so that the PSF does not vary with depth. In addition, we assume that there is no wavelength variation in the PSF as demonstrated in \cite{monakhova2020spectral}. 

\subsection{2D imaging with erasures}
For the case of 2D imaging with erasures, we multiply the result of Eq. \ref{eq:forward_2D} with a binary mask, denoted $\mathbf M[x,y]$, that zeros out a subset of the sensor pixels to model the erasure pattern:
\begin{align}
\label{eq:forward_2D_erasures}
 \mathbf{b}[x,y] &= \mathbf M[x,y] \cdot \text{crop} \Big( \mathbf v[x,y] * \mathbf h[x,y]) \Big)\\
 &= \mathbf A_{\text{2D erasures}} \mathbf v.
\end{align}
As we show in Sec. \ref{sec:results}, compressed sensing enables recovery of a full image, even in the presence of a significant fraction of dead or missing pixels. Although commercial sensors are screened to minimize dead pixels, this scenario is useful for exploring the limits of compressive sensing in lensless cameras since we can synthetically increase the percentage of erasures to increase the ratio of data reconstructed to data measured.

\subsection{Higher dimensions: video and hyperspectral}
Rather than sampling only a subset of pixels in order to reconstruct a 2D image, compressed sensing can instead be used to reconstruct 3D datasets from fully-sampled 2D images. If the higher-dimensional data is physically encoded into different subsets of the pixels, one can use Eq. \ref{eq:forward_2D_erasures} to recover a collection of 2D images from a single acquisition. As described previously, our single-shot video example uses the rolling shutter to encode temporal information into different pixels, and our hyperspectral example uses a spectral filter array to encode wavelength information. We denote this extra dimension (time/wavelength) generically as the $k$-dimension. Sequential recovery using Eq. \ref{eq:forward_2D_erasures} prevents incorporating priors along the $k$-dimension, so we use the following model that depends on the full datacube:
\begin{align}
\label{eqn:forward_video}
 \mathbf b &= \sum_{k=0}^{N_k} \mathbf M_{k}[x,y] \cdot \text{crop} \Big( \mathbf v[x,y,k] {*} \mathbf h[x,y] \Big) \\
 &= \mathbf A_{k\text{-D}} \mathbf v.
\end{align}
Here, $N_k$ is the number of discrete points along the $k$-dimension, and $\mathbf{M}_{k}[x,y]$ is a masking function, which depends on $k$, and selects the sensor pixels corresponding to each video frame/wavelength. The convolution, $*$ is only over the two spatial dimensions.

For the high-speed video case $\mathbf{M}_{k}[x,y]$, referred to as the shutter function, is based on the rolling shutter. Sensors typically have either a single shutter or a dual shutter, which we use in this work.  For a single shutter, the sensor reads the pixels one horizontal line at a time, moving from the top to the bottom of the sensor; for a dual-shutter, the sensor reads pixels from two horizontal lines that move from the top and bottom of the sensor towards the middle, Fig.~\ref{fig:forward}(b). 

For the hyperspectral case $\mathbf{M}_{k}[x,y]$, referred to as the filter function, is determined by the spectral filter array. Each filter pixel acts as narrow-band spectral filter which integrates light within a certain wavelength range and blocks out light at other wavelengths. We approximate this as a finite sum across spectral bands with a non-binary filter function.

\section{Inverse problem}
Given our sensor measurement $\mathbf b$, our goal is to recover the scene $\mathbf v$. For the 2D erasures scenario $\mathbf v$ is a 2D image, whereas for single-shot video $\mathbf v$ is a video consisting of two spatial and one temporal dimension, and for single-shot hyperspectral $\mathbf v$ is a hyperspectral volume consisting of two spatial and one spectral dimension. In all cases, the problem is underdetermined, since we aim to recover more than we measure. First, we describe the traditional reconstruction methods based on convex optimization which will serve as our baseline comparison, and then we describe our untrained network reconstruction method.


\begin{figure*}[t]
\includegraphics[width=\linewidth]{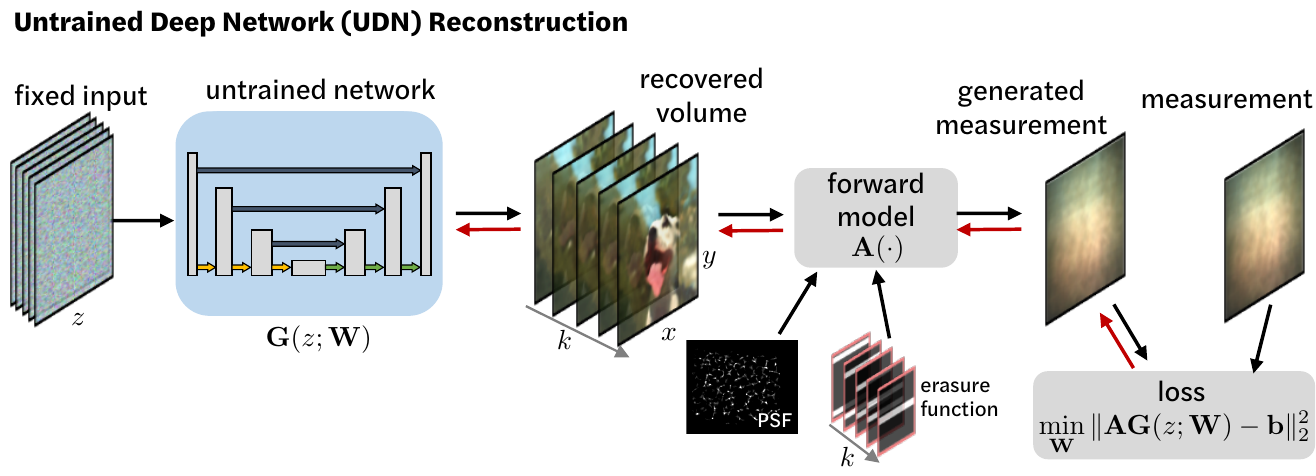}
   \caption{\textbf{Overview of our Untrained Deep Network (UDN) reconstruction pipeline}. An untrained network with randomly initialized weights takes in a fixed, random input vector. The network outputs a sequence of images along the $k$-axis (for video imaging, one image for each time point), and we pass this output volume into our known imaging forward model (defined by the PSF calibration image and known erasure function) to generate a single simulated measurement.  We compare the generated measurement with our captured measurement and use the difference between the two to update the untrained network parameters.}
\label{fig:overview}
\end{figure*}

\subsection{Traditional inverse problem}
Due to incoherence of the measurement system that comes from our use of a multiplexing phase optic (diffuser), compressive sensing can be utilized to recover the image. By compressive sensing theory, we can formulate our inverse problem as follows:

\begin{equation}
\label{eqn:inverse1}
 \hat{\mathbf v} = \arg\min_{\mathbf v\geq 0} \frac{1}{2} \| \mathbf b - \mathbf A \mathbf v \|_2^2 + \tau R(\mathbf v), 
\end{equation}
\noindent where $R(\mathbf v)$ is a prior on the scene, and is often of the form $\| \mathbf D \mathbf v \|_1$, where $\mathbf D$ is a sparsifying transform. $\mathbf A$ is our forward model, which can be of the form $\mathbf A_{\text{2D}}$, $\mathbf A_{\text{2D erasures}}$, or $\mathbf A_{\text{k-D}}$.  

In practice, 2D or 3D TV priors work well for a variety of scenes, and are implemented by defining the regularizer term as $R(\mathbf v) = \| \nabla_{xyk} \mathbf v\|_1 $, where $\nabla_{xyk} = [\nabla_x \nabla_y \nabla_k]^T$ is the matrix of forward finite differences in the $x$, $y$, and $k$ directions. Convex optimization approaches such as fast iterative shrinkage-thresholding algorithm (FISTA)~\cite{beck2009fast} or alternating direction method of multipliers (ADMM)~\cite{boyd2011distributed} can be used to efficiently solve this problem.  In each case, the prior is hand-tuned for the given application by varying the amount of regularization through the tuning parameter $\tau$, which trades off data-fidelity and regularization. As a baseline comparison, we use FISTA with 2DTV for our 2D imaging with erasures problem and weighted anisotropic 3DTV for our single-shot video and single-shot hyperspectral imaging~\cite{kamilov2016parallel}. We include an additional comparison using PnP-BM3D, PnP-BM4D~\cite{maggioni2012nonlocal}, and a pretrained PnP denoiser network~\cite{zhang2020plug} in Supplement 1. 

\subsection{Untrained network}
In this work, we propose to instead use an untrained network for the image reconstruction. It has been shown that neural networks are good at representing and generating natural images; thus, minimizing an image generator network instead of directly solving for the image could be a good way to effectively regularize with a custom deep prior. 

Rather than solving for the image $\mathbf v$ directly as before, we solve for the image indirectly as the output of the generator network $G(z; \mathbf W)$.  This output image $\mathbf v_{gen}$ is then passed through our imaging forward model $\mathbf A$ and compared against the measured image $\mathbf b$. The network weights are updated based on the difference between the generated measurement and the true measurement, which corresponds to solving the following problem: 

\begin{align}
    \mathbf W* =& \arg\min_{\mathbf W} \frac{1}{2} \| \mathbf b - \mathbf A \mathbf G (z;\mathbf W) \|_2^2 \\
    =& \arg\min_{\mathbf W} \frac{1}{2} \| \mathbf b - \mathbf A \mathbf v_{gen} \|_2^2,
\end{align}

\noindent where our network $G(z; \mathbf W)$ has a fixed input $z$ and randomly initialized weights $\mathbf W$. The network output $\mathbf v_{gen}$ is the reconstructed image. $z$ can be thought of as a latent code to our generator network and is randomly initialized and held constant.  We update the weights of the network via backpropogation in order to minimize the loss between the actual measurement and the generated measurement (Fig.~\ref{fig:overview}). This process must be repeated for each image reconstruction, since there are no `training' and `testing' phases as there are for deep learning methods with labeled data.

As in~\cite{ulyanov2018deep}, we utilize an encoder-decoder architecture with skip connections and keep the input to the network fixed. See Supplement 1 for details on our network architecture and hyperparameters for each experiment.


\section{Implementation details}
For the 2D erasures case, we utilize an existing 2D lensless imaging dataset which consists of pairs of 2D lensless camera measurements and corresponding lensed camera ground truth images~\cite{monakhova2019learned}. This dataset is advantageous because it includes experimental lensless measurements with real sensor noise and other non-idealities, but also provides labeled ground truth data from an aligned lensed camera. For our simulation results, we utilize the ground truth images from the dataset to generate simulated measurements using the forward model in Eq. \ref{eq:forward_2D_erasures}; for our experimental results, we directly use the lensless camera measurements. For both, we synthetically add erasures to the measurement by point-wise multiplying it with an erasure function with 0\%, 50\%, 90\%, 95\%, and 99\% erasures, picking a random subset of indices to erase. 

For single-shot video and single-shot hyperspectral imaging, there does not exist experimental data with associated ground truth. For our experimental data analysis, we utilize experimental raw sensor measurements from~\cite{antipa2019video} and~\cite{monakhova2020spectral}. This includes measurements taken from a PCO Edge 5.5 sCMOS camera with a dual rolling shutter and a homemade random diffuser (for single-shot video), and measurements from a hyperspectral DiffuserCam with a spectral filter array and Luminit diffuser~\cite{luminit_diffuser} (for hyperspectral imaging). For our simulations, we use the PSF and shutter/filter functions from these systems to simulate our sensor measurements. 

We implement our network and differentiable forward model with mean square error (MSE) loss in Pytorch and run our reconstructions on a Titan X GPU with the ADAM optimizer throughout training. We perform early stopping to obtain the best reconstructions, as described in~\cite{ulyanov2018deep}, with reconstructions ranging from 1,000 - 100,000 iterations, which generally takes several hours. This process must be completed for every new image; unlike with deep learning methods that use training data, there is no distinction between training and testing and instead the network parameters must be re-optimized for every reconstruction. For reproducibility, training code will be available here upon publication of the paper: \textit{https://github.com/Waller-Lab/UDN/} .

\begin{figure*}[t]
\includegraphics[width=\linewidth]{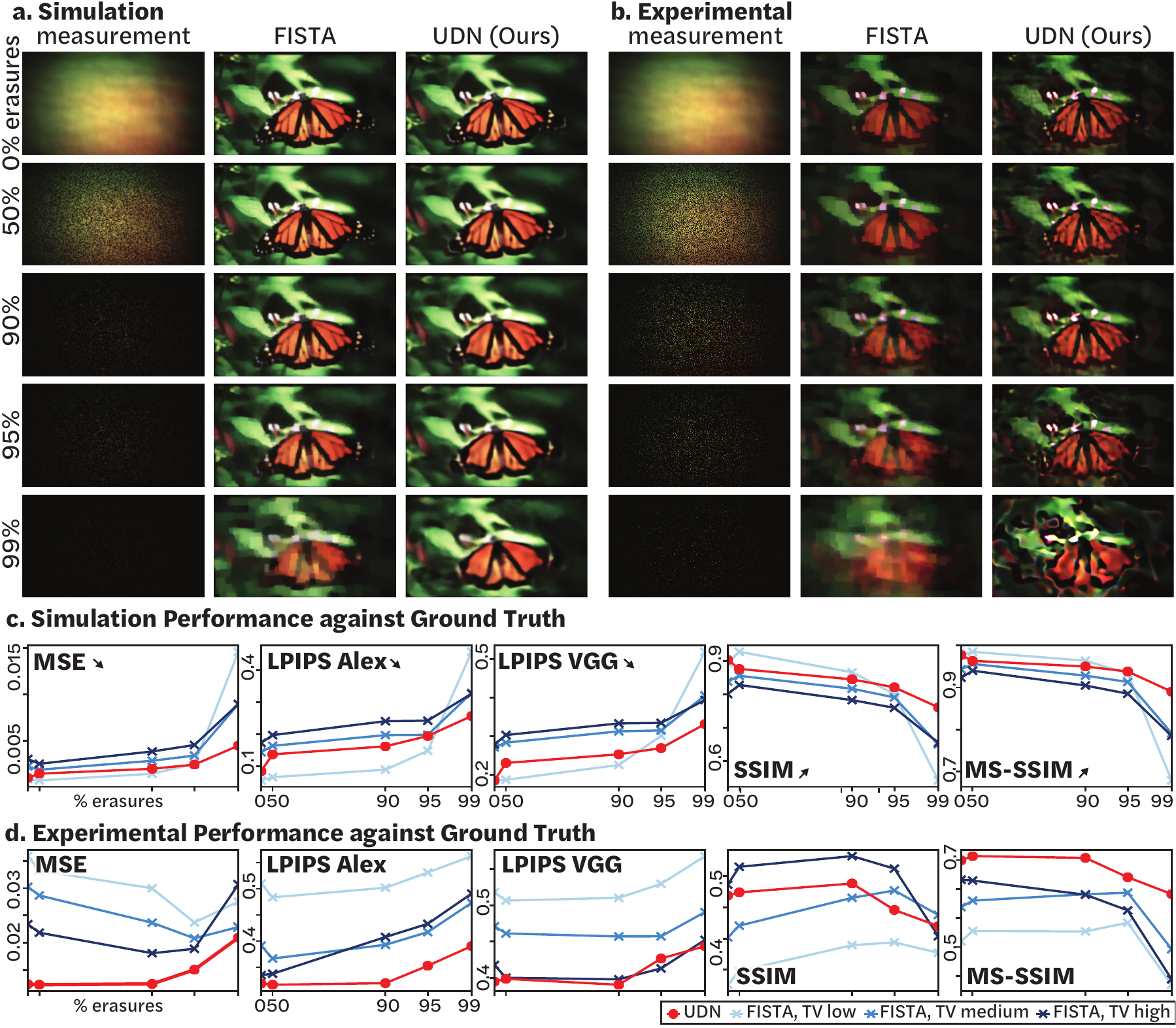}
   \caption{\textbf{2D Compressive imaging with erasures: simulation and experimental results}. Reconstruction results for increasing numbers of pixel erasures, showing the measurement along with the best fast iterative shrinkage-thresholding algorithm (FISTA) reconstruction (based on the LPIPS Alex metric) and the UDN reconstruction for (a) simulated measurements and (b) experimental measurements with synthetically added erasures. (c) and (d) compare our performance against the ground truth for several quality metrics (arrows indicate which direction is better), showing that in simulation UDN outperforms FISTA only at 99\% erasures, whereas in experiments with real noise and non-idealities, UDN outperforms FISTA on the perceptual image metrics, LPIPS Alex and MS-SSIM, as well as on MSE.}
\label{fig:2D_images}
\end{figure*}

\begin{figure*}[t!]
\includegraphics[width=\linewidth]{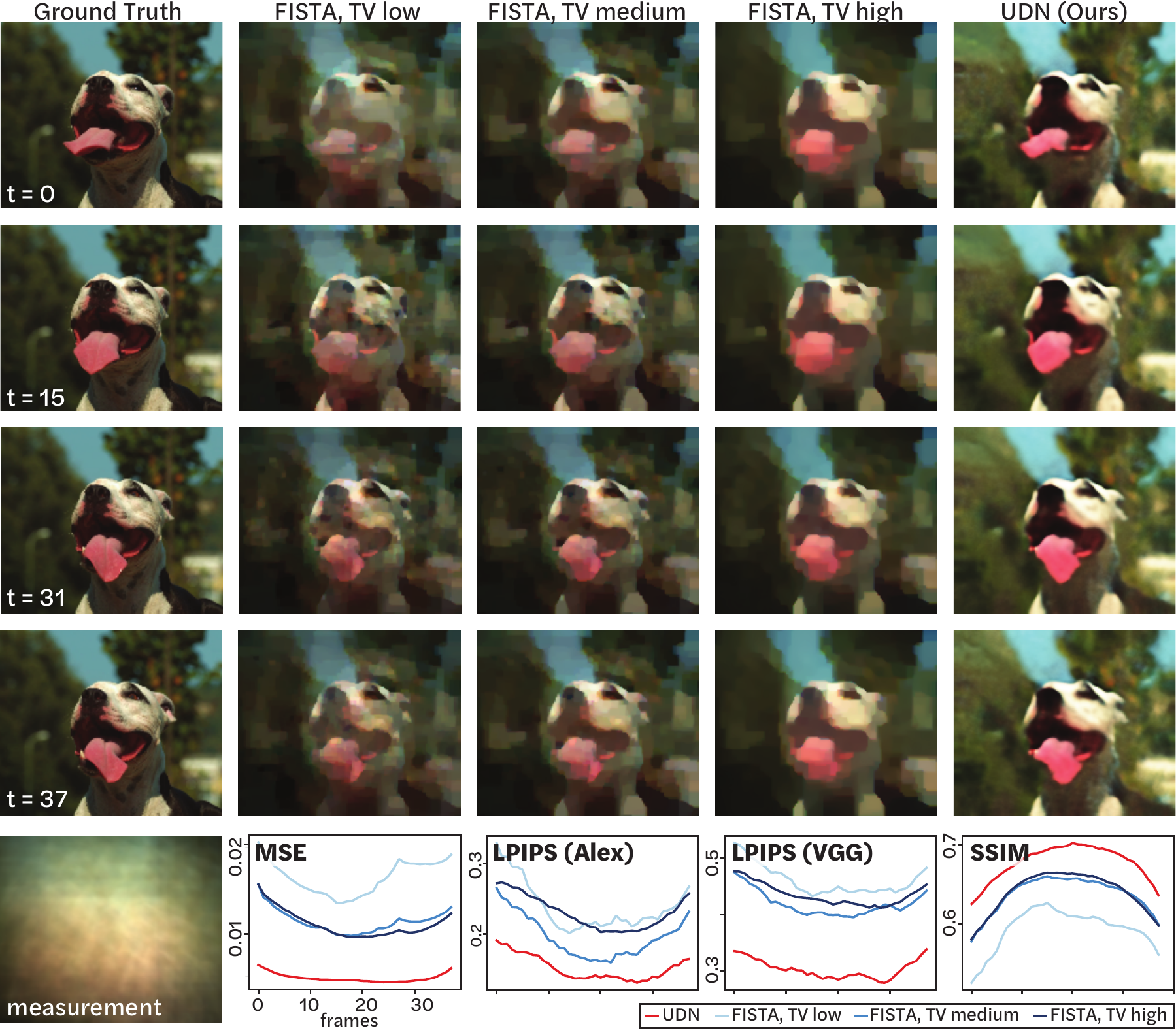}
   \caption{\textbf{Simulation results on single-shot video}. We recover a 38-frame video from a single measurement (bottom left).  We show four sample reconstructed video frames in the top four rows, and plot the quality metrics for all frames below. Here we see that our reconstruction has sharper features across frames, enabling superior recovery especially for the first and last frames, where FISTA has more pronounced reconstruction artifacts. We achieve better MSE, LPIPS, and SSIM scores across frames (bottom right). See Visualization 1 for the full video.}
\label{fig:video_dog}
\end{figure*}


\section{Results} \label{sec:results}
We compare the results of our untrained reconstruction method against the standard FISTA reconstruction with TV regularization. For all three cases, we present both simulation results as well as experimental validation. For the FISTA reconstructions, we reconstructed images with three different amounts of TV regularization, since more regularization leads to improved image quality and lower mean squared error, but tends to blur high-frequency information. Less regularization reveals high-frequency information, but at the price of increased reconstruction artifacts. When ground truth is available (in all simulations and in experiment for 2D erasures), we compare our reconstructed images to the ground truth via a variety of quality metrics: mean squared error (MSE), learned perceptual similarity metric (LPIPS) based on AlexNet and VGG~\cite{zhang2018unreasonable}, the structural similarity index measure (SSIM), and the multi-scale structural similarity measure (MS-SSIM)~\cite{wang2003multiscale}.  MSE is a standard image metric, but tends to favor low-frequency information. SSIM and MS-SSIM are both perception-based metrics, with MS-SSIM using multiple image scales and achieving better perceptual performance than SSIM. Both LPIPS metrics are learned perpetual similarity metrics based on deep networks, with LPIPS VGG being closer to a traditional perceptual similarly metric. Each method has its strengths and weaknesses, and therefore we provide comparisons using each of the metrics when possible (MS-SSIM requires a certain minimum image size which precludes us from using it for the single-shot video and single-shot hyperspectral cases, and LPIPS only works on RGB color images, so we cannot use it for single-shot hyperspectral). For MSE, LPIPS Alex, and LPIPS VGG, a lower score is better, whereas for SSIM and MS-SSIM, a higher score is better.

\subsection{2D compressive imaging}
\subsubsection{Simulation}
First, we simulate a noise-free 2D measurement with increasing numbers of randomly-distributed pixel erasures  (0\%, 50\%, 90\%, 95\%, and 99\%), then compare reconstruction results using both FISTA and our UDN (Fig.~\ref{fig:2D_images}). FISTA uses high, medium, and low amounts of TV corresponding to $\tau$= 1e-4, 5.5e-4, 1e-5, respectively.

We compare our performance against the ground truth images on our five image quality metrics for an increasing percentage of erasures in Fig.~\ref{fig:2D_images}(b) and show the corresponding images in Fig.~\ref{fig:2D_images}(a) for our method vs. the best FISTA result (based on the LPIPS-Alex metric). From this, we can see that FISTA and our method perform similarly well for 0\%-95\% erasures, but our method has improved performance for 99\% erasures. This shows that TV regularization is a sufficient prior for smaller numbers of erasures in the absence of noise or model non-idealities, but as the problem becomes severely underdetermined, our UDN provides a better image prior than TV and leads to improved image quality on all of our performance metrics.

\subsubsection{Experimental}
Using experimentally captured images from~\cite{monakhova2019learned}, we synthetically add random pixel erasures to the measured data and compare our reconstructions against the ground truth lensed images using the five image metrics described above. As a baseline, we compare against FISTA with high, medium, and low amounts of TV ($\tau$= 1e-1, 1e-2, 1e-3), shown in Fig.~\ref{fig:2D_images}(d).  A visual comparison with the best FISTA reconstruction (based on the LPIPS-Alex metric) is shown in Fig.~\ref{fig:2D_images}(c). Unlike in the noise-free simulation, our method consistently performs better than FISTA for all amounts of erasures on the MSE, LPIPS Alex, and MS-SSIM metric, resulting in a clearer and sharper reconstruction. Thus, our method gives better performance improvements over FISTA for real-world datasets, likely because the UDN provides a better image prior and outperforms TV in the presence of real measurement noise and imaging non-idealities.

TV is a very commonly used prior that is computationally efficient, but other priors can be incorporated using the PnP framework. See Supplement 1 for additional comparisons against PnP-BM3D and a PnP pretrained denoiser network, demonstrating that the UDN outperforms these methods as well. 

\begin{figure*}[t]
\includegraphics[width=\linewidth]{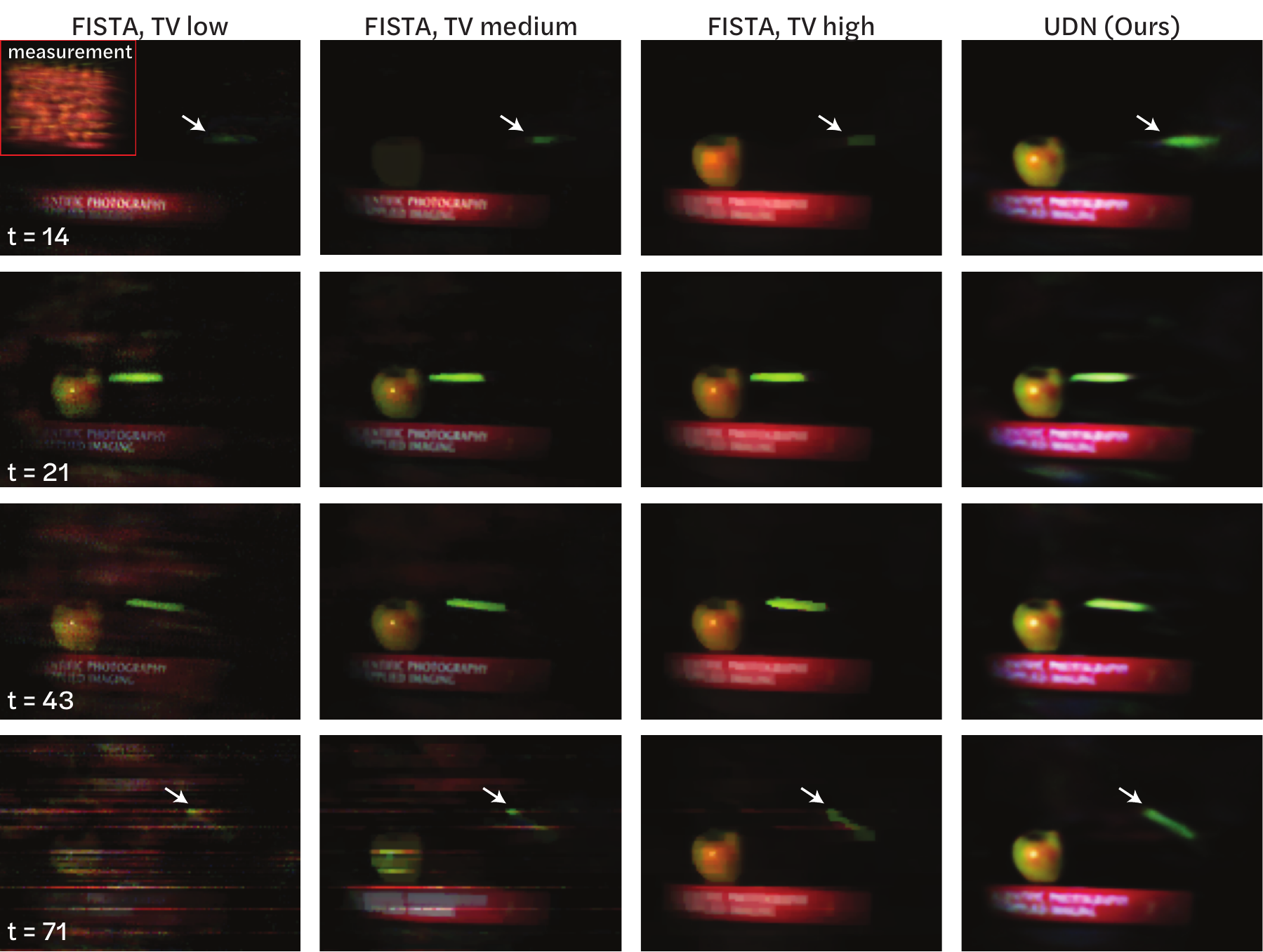}
   \caption{\textbf{Experimental results on single-shot video}. We recover 72 frames from a single measurement (top left). The rows show four sample reconstructed frames from our 72-frame reconstruction, with both our UDN method and FISTA with three different amounts of TV. Here we see that our reconstruction has sharper features across frames and better captures motion within the video. See Visualization 2 for the full video and Visualization 3 for a second experimental example.}
\label{fig:video_dart}
\end{figure*}

\begin{figure*}[t]
\includegraphics[width=\linewidth]{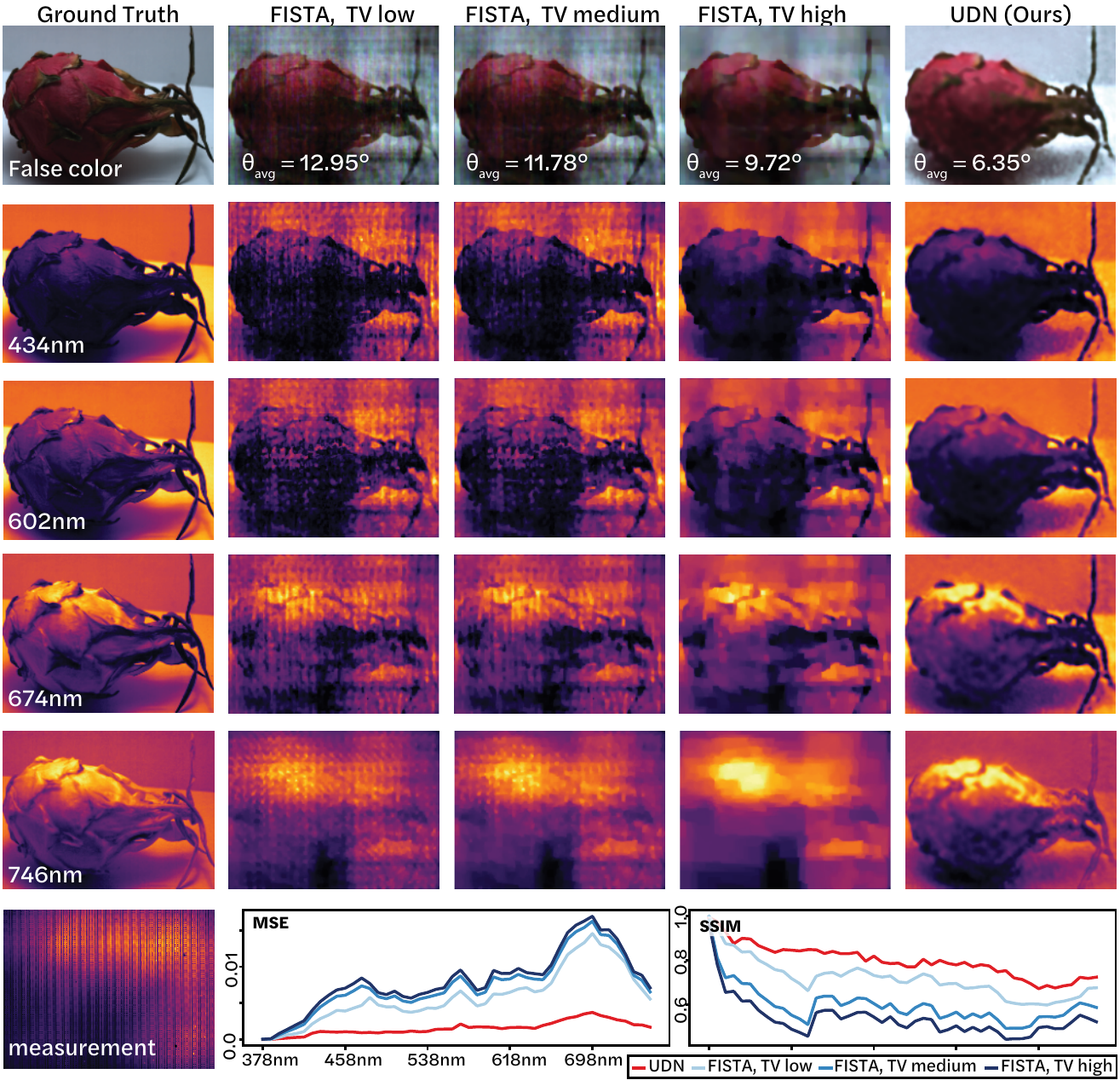}
   \caption{\textbf{Simulation results on hyperspectral data}. We display a false-color image of the recovered 64-channel hyperspectral volume (top row), along with four selected spectral slices. The quality metrics are plotted for each wavelength at bottom right. Here we see that our reconstruction has sharper features and fewer artifacts than FISTA, and achieves a better MSE and SSIM score across all wavelengths. In addition, UDN achieves better cosine similarity ($\theta_{\text{avg}}$) to the ground truth spectra than FISTA. See Visualization 4 for the full hyperspectral volume.}
\label{fig:spectral_fruit}
\end{figure*}

\begin{figure}[t]
\includegraphics[width=\linewidth]{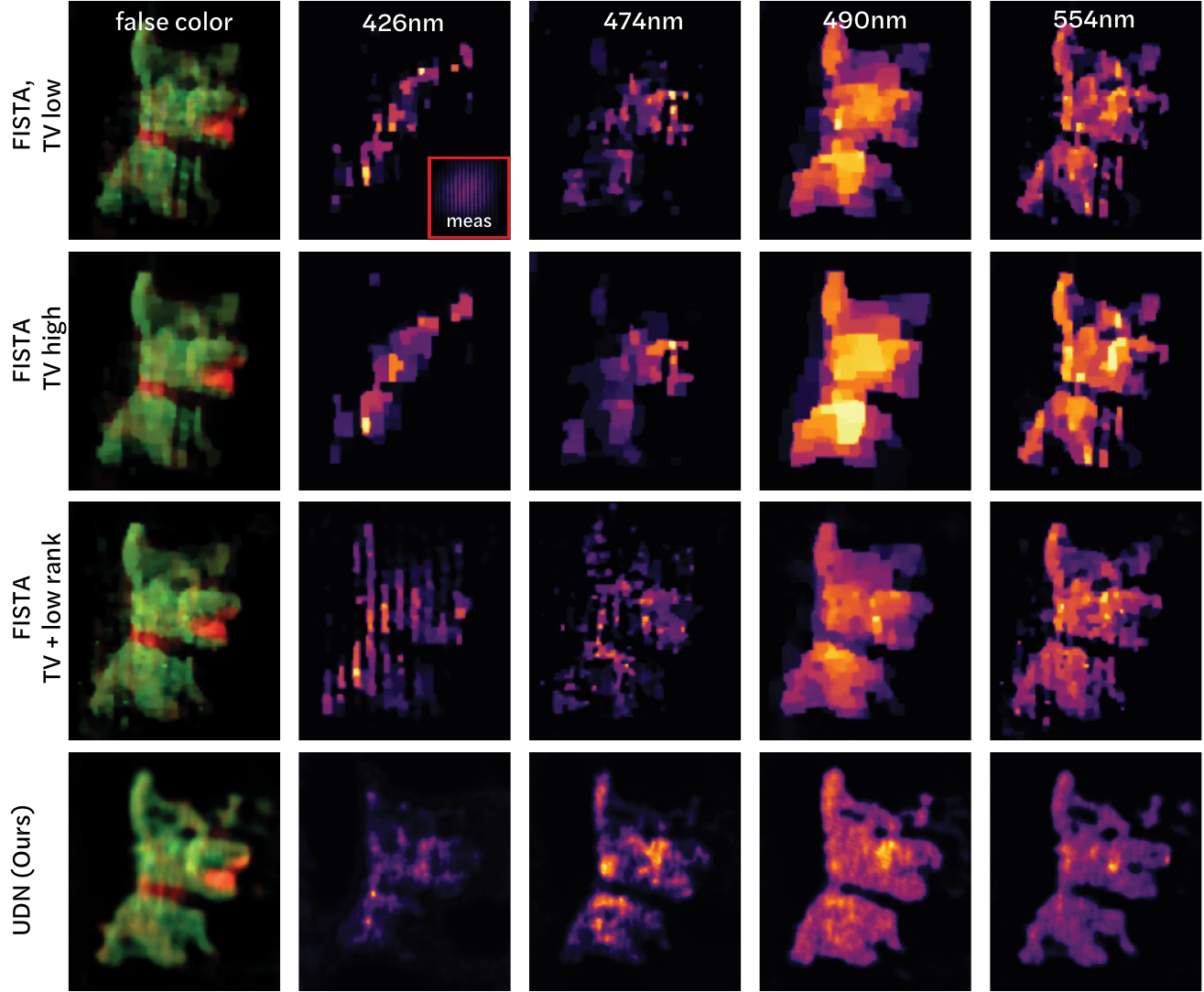}
   \caption{\textbf{Experimental results on hyperspectral data}. We show a false-color image of the recovered 32-channel hyperspectral volume (top row), along with four spectral slices. Here we see that our reconstruction has sharper features and fewer artifacts across wavelengths. See Visualization 5 for the full hyperspectral volume.}
\label{fig:spectral_dog}
\end{figure}

\subsection{Single-shot Video}
\subsubsection{Simulation}
Using an experimentally captured PSF and shutter function from~\cite{antipa2019video}, we simulate a compressive video measurement, shown in Fig.~\ref{fig:video_dog} top left, using a 38-frame video. Fig. ~\ref{fig:video_dog} shows the results of our reconstruction compared to the FISTA result for several frames of the video (with $\tau$ = 1e-2, 1e-3, 5e-4 for FISTA). It is evident that our method generates a more visually appealing result with fewer artifacts, while preserving high-resolution features throughout the frames. This is quantified in Fig.~\ref{fig:video_dog}(bottom), showing that UDN has a significantly better performance on all metrics. For each of the FISTA results, the values are worse at the beginning and end of the video sequence, which is consistent with~\cite{antipa2019video}. Our method similarly has worse performance for the first and last frames, but the difference is not as pronounced, resulting in more uniform image quality throughout the recovered video. 

\subsubsection{Experimental}
Next, we recover 72-frame videos from a single experimental measurement, Fig.~\ref{fig:video_dart}. When compared to the FISTA reconstructions (with $\tau$ = 1e-2, 1e-3, 1e-4), we can see that our method has more uniform image quality throughout. The foam dart has significant artifacts in the FISTA reconstruction for the first and last scenes of the video, even disappearing for low TV. Our method appears to capture the dart motion throughout the video frames and has fewer noticeable artifacts than the FISTA reconstruction. There is no ground truth available for this data, but our method seems to produce a more realistic and visually appealing reconstruction than FISTA in terms of the dart motion. FISTA with smaller amounts of TV is able to better recover the text on the book in the scene, but at the expense of reduced image quality and more artifacts throughout the video. Our method is not quite able to recover the book text, but has more uniform image quality and better captures the motion in the scene.

\subsection{Single-shot hyperspectral}
\subsubsection{Simulation}
Finally, we recover a hyperspectral volume with 64 wavelengths from a single simulated measurement. We simulate the measurement using an experimentally captured PSF and filter function from~\cite{monakhova2020spectral}, along with a ground truth hyperspectral image from~\cite{ennis2018hyperspectral} with 64 spectral channels. Figure~\ref{fig:spectral_fruit} shows the results of our reconstruction compared to FISTA with high, medium, and low amounts of TV ($\tau$= 3e-7, 6e-7, and 3e-6). We can see that UDN preserves more features across wavelengths and has fewer artifacts than FISTA. We compare the MSE and SSIM across the methods, Fig.~\ref{fig:spectral_fruit}(bottom). We can see that UDN has significantly better MSE and SSIM values than FISTA across wavelengths. In addition, we report the average cosine distance between the spectral profiles in the reconstruction and ground truth images (Fig.~\ref{fig:spectral_fruit}). The cosine distance is especially important for hyperspectral imaging due to its role in hyperspectral classification. UDN provides a lower average cosine distance than FISTA, indicating that UDN achieves a better recovery of the spectral profiles. We note that FISTA with high TV achieves a better cosine distance than FISTA with low TV, but at the price of worse spatial resolution. Our UDN method achieves both better spatial quality (MSE and SSIM) as well as better spectral performance (cosine distance) than FISTA. 

\subsubsection{Experimental}
We test our performance on an experimentally captured measurement~\cite{monakhova2020spectral} of a Thorlabs plush dog illuminated by a broadband lightsource, recovering 32 spectral channels from a single measurement (downsampled 2$\times$ spectrally). We compare against FISTA with low and high TV ($\tau$= 3e-4 and 5e-5) along with the best reconstruction from~\cite{monakhova2020spectral} (FISTA TV + low rank), down-sized to match our reconstruction size. While ground truth images do not exist for this data, UDN appears to provide more consistent image quality and retains more of the details across wavelengths than FISTA, Fig.~\ref{fig:spectral_dog}. 

\section{Discussion}
Untrained networks offer a number of distinct advantages for compressive lensless imaging. First, they do not require any training data, as opposed to deep learning-based methods. Training data is especially hard or impossible to acquire for higher-dimensional imaging, such as for high-speed video, hyperspectral, or 3D imaging, so this feature is particularly useful. Untrained networks can serve as a better prior for certain high-dimensional imaging problems off-the-shelf, potentially enabling better reconstructions for a number of compressive imaging modalities. 

Currently, the main limitations of untrained networks are: 1) memory-constraints and 2) speed. First, many of our reconstructions are GPU memory limited. To take advantage of accelerated GPU processing, the entire untrained network must fit in memory on the GPU, limiting the size of reconstructions that we can process, and we find that we can process much larger images and volumes with FISTA than we can with UDN. In this work, our single-shot video and single-shot hyperspectral measurements are downsized between 2-16$\times$ from the original size in order to fit in the GPU, limiting our resolution. Looking forward, larger GPU sizes or clever computational techniques to better utilize GPUs for memory-limited problems could improve this and enable the reconstruction of larger images and volumes. Next, the speed of the untrained network reconstructions is generally an order-of-magnitude slower than standard FISTA reconstructions. Thus, our method is best suited for applications where a real-time reconstruction is not needed, since typical untrained reconstructions take between 1-5 hours. As machine learning speeds and processors improve, we expect these factors to be less limiting. 

Given the benefits and limitations of untrained networks, we envision this reconstruction method to be useful for a certain subset of problem where it is difficult or impossible to obtain ground truth data for training, but where there are little or no time constraints on the reconstruction. Given the fact that no training data is needed, this method can be applied to many different imaging problems without needing to collect new datasets. UDNs are especially promising for imaging dense, natural scenes where TV can be a poor prior. 


One interesting open problem for untrained reconstructions is the network choice. In our experiments, we utilized a convolutional network with an encoder-decoder structure and found that this worked well; however, our network tended to blur out high frequency features and there may be other architectures that could potentially provide better priors. For instance, deep decoder~\cite{heckel2018deep} has been demonstrated for similar tasks, but we found that for our application it required more iterations to converge and was outperformed by an encoder-decoder structure, demonstrating that the choice of network architecture is important and application-specific. Although our network worked well for photographic scenes, other network architectures may be more appropriate for other types of images, such as fluorescent biological targets which may have very different statistics than photographic scenes. 

\section{Conclusion}
We have demonstrated that untrained networks can improve the image quality for lensless compressive imaging systems. We tested untrained networks on 2D lensless imaging with varying amounts of erasures, and we demonstrated their effectiveness on single-shot compressive video and single-shot hyperspectral imaging, in which we recover a full 72-frame video or 32 to 64 spectral slices, respectively, from a single measurement. In each case, we showed both in simulation and experiment that untrained networks can have better image quality than compressive-sensing-based minimization using total-variation regularization, demonstrating that non-linear networks can be a better prior than TV for dense, natural scenes. We believe that untrained networks are especially promising for situations in which training data is difficult or impossible to obtain, providing a better imaging prior for underdetermined reconstructions. 

\begin{backmatter}
\bmsection{Funding}
Placeholder

\bmsection{Acknowledgments}
Vi Tran acknowledges support from the Transfer-to-Excellence REU program, funded by the Hopper Dean Foundation and hosted by the Center for Energy Efficient Electronics Science (NSF Award 0939514). 
\bmsection{Disclosures}
The authors declare no conflicts of interest

\bmsection{Data Availability Statement}
All data will be available here:  https://github.com/Waller-Lab/UDN/ upon publication of the paper.

\bmsection{Supplemental document}
See Supplement 1 for supporting content. 

\end{backmatter}

\bibliography{sample}

\end{document}